\documentclass[aps,prl,twocolumn,nopreprintnumbers,noeprint]{revtex4-1}
\usepackage{tikz}
\usetikzlibrary{arrows}
\usepackage{graphicx}
\usepackage{dcolumn}
\usepackage{bm}
\usepackage{amsmath}
\usepackage{amsfonts}
\usepackage{psfrag}
\usepackage{graphicx}
\usepackage{color} 

\def\bra#1{\mathinner{\langle{#1}|}} 
\def\ket#1{\mathinner{|{#1}\rangle}}

\newcommand{\Eq}[1]{Eq. (\ref{#1})}

\begin{document}

\title{Electro-optical sampling of quantum vacuum fluctuations in dispersive dielectrics}

\author{Simone \surname{De Liberato}}
\affiliation{School of Physics and Astronomy, University of Southampton, Southampton, SO17 1BJ, United Kingdom}

\begin{abstract}
Electro-optical sampling has been recently used to perform spectrally-resolved measurements of electromagnetic vacuum fluctuations. In order to understand which information on the ground state of an interacting system can be acquired thanks to this technique, in this paper we will develop the quantum theory of electro-optical sampling in arbitrary dispersive dielectrics. Our theory shows that a measure of the  time correlations of the vacuum fluctuations  effectively implements an ellipsometry measurement on the quantum vacuum, allowing to access the frequency-dependent dielectric function. We discuss consequences of these results on the possibility to use electro-optical sampling to probe the population of ground-state virtual photons in the ultrastrong light-matter coupling regime. 
\end{abstract}

\maketitle

\section{Introduction}

The Heisenberg uncertainty principle constrains an oscillator in its ground state to have a finite kinetic energy. In the context of quantum electrodynamics this leads to the picture of an empty space populated by random fluctuations of quantum nature. 
The effect of quantum vacuum fluctuations (QVF) can be most easily recognised in any spontaneous radiation process. Sending excited atoms flying in a photonic cavity and using sub-wavelength imaging to pinpoint the location of photon emission, the spatial distribution of QVF was thus directly measured \cite{Lee2014}. A different approach relies on the nonlinear effect QVF can have upon light propagating into a medium. A detection scheme based on electro-optical sampling has been successfully used to measure both the intensity of the electric field in the vacuum \cite{Moskalenko2015,Riek2015,Riek2017,Riek2017a} and
its time- and space-dependent correlation function \cite{Benea-Chelmusarxiv}.

Such a technique could {\it a priori} reveal itself a useful tool to investigate the ground state properties of interacting systems, but to which point it can be used to probe the structure of the quantum vacuum is presently unknown. 
In particular in Ref. \cite{Benea-Chelmusarxiv} it is suggested that spectrally-resolved electro-optical sampling of QVF could provide a first direct evidence of the presence of virtual photons in the ground state of a system in the ultrastrong light-matter coupling regime \cite{Kockum2019,FornDiaz2018}. In this regime the strength of the light-matter interaction is large enough to hybridise the uncoupled electromagnetic vacuum $\ket{0}$ with excited states, leading to a novel coupled polaritonic ground state $\ket{P}$. The form of such a coupled ground state was initially calculated analytically in Ref. \cite{Quattropani1986}, showing it has the form of a two-modes squeezed vacuum, containing a population of virtual photons. Those virtual photons, localised in proximity of the quantum emitter \cite{Peropadre2013a,Munoz2018}, can become real and be radiated when the system parameters are modulated in time \cite{DeLiberato2007,Dodonov2008,Carusotto2012,Huang2014,Hagenmuller2016,DiStefano2017}, an effect reminiscent of the Dynamical Casimir effect \cite{Wilson2011}. 
Notwithstanding a remarkable interest, both theoretical \cite{Ridolfo2013,Garziano2015,Schafer2018,FalciarXiv} and experimental \cite{Anappara2009,Todorov2010,Scalari2013,Gambino2014,Bayer2017,Li2018,Yoshihara2018} in the physics and phenomenology of the ultrastrong coupling regime, for the moment no direct evidence of the virtual photons has been obtained.

In order to clarify which features of the quantum vacuum can be measured using electro-optical sampling, and in particular if we can use it to directly measure ground state virtual photons, in this paper we will develop the quantum theory of spectrally-resolved electro-optical sampling of QVF in dispersive linear materials. 
Using such a theory we will be able to demonstrate that the time-dependent correlation function of the QVF, once normalised over the free-space vacuum value, provides access to the spectrally-resolved dielectric function.  
On one hand this implies such a technique can be used to perform ellipsometry characterisation of linear optical properties without the need of a resonant probe beam. On the other hand, the fact that all the quantities accessible with such a measurement can generally also be accessed by linear-optical techniques, raises doubts on the possibility of using it as a direct test for the presence of virtual photons.
\begin{figure}[htbp]
\begin{center}
\includegraphics[width=6cm]{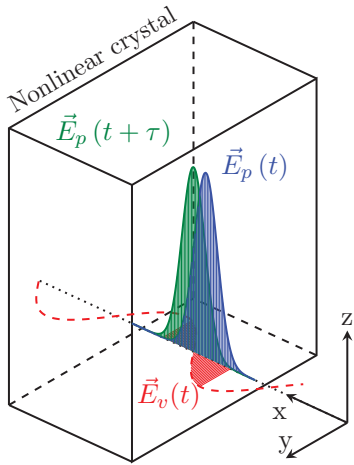}
\caption{Sketch of electro-optical sampling measurement. Two linearly polarised sub-cycle probe pulses $\vec{E}_p(t)$ propagate in a nonlinear crystal with a  delay $\tau$. Due to the nonlinearity each pulse interacts with the QVF of the field in the orthogonal polarization $\vec{E}_{v}(t)$.} \label{Fig_Sketch}
\end{center}
\end{figure}

\section{Electro-optical sampling of QVF}

Electro-optical sampling consists in mixing an intense, linearly polarised, sub-cycle probe pulse with a weak field perpendicular to it, allowing to observe the rotation induced by the weak field on the polarization of the probe. Fixing an orthogonal axis system as in Fig. \ref{Fig_Sketch}, in which $x$ is the direction of beam propagation and the probe is polarised along $z$, and considering the nonlinear crystal to be properly oriented, the quantity measured is then the $y$ component of the electric field. The operator corresponding to such a measurement is the electro-optical operator  \cite{Moskalenko2015}.
\begin{align}
\label{Seo}
\hat{S}_{eo}(t)&=\sum_{k}\sqrt{\frac{C\hbar\Omega_k}{2\epsilon_0\epsilon_rV}}
\left[\hat{a}_{k}R(\Omega_k)e^{-i\Omega_k t}-\text{h.c.}\right],
\end{align}
where the sum is over all the $y$-polarised paraxial modes of wavevector $k$ and frequency $\Omega_k=\frac{ck}{\sqrt{\epsilon_r}}$, with annihilation operator $\hat{a}_{k}$, and $R(\Omega)$ is a low-pass filter, dependent on the phase-matching condition of the nonlinear process and proportional to the spectral autocorrelation function of the probe beam.  The relative dielectric permittivity of the nonlinear crystal is $\epsilon_r$ and h.c. stands for Hermitic conjugate. 
The volume $V=LS$ in \Eq{Seo} depends on the transversal surface of the probe beam ($S$) and the paraxial quantisation length ($L$), and $C$ is a function depending both on the probe beam and on the electro-optical crystal used.
In the ground state the expectation value of $\hat{S}_{eo}(t)$ vanishes, and information has thus to be extracted by its higher order momenta. If the measure is repeated after a short delay $\tau$, such a technique can then give us access to the time-dependent correlation function
\begin{eqnarray}
\hat{G}_{eo}(\tau)=-\frac{1}{2C} \left\{\hat{S}_{eo}(t+\tau),\hat{S}_{eo}(t)\right\},
\end{eqnarray}
where $\{\cdot,\cdot\}$ indicates the anticommutator. Its expectation value in the electromagnetic vacuum $\hat{a}_{k}\ket{0}=0$ reads
\begin{eqnarray}
\label{Geot}
\bra{0} \hat{G}_{eo}(\tau)\ket{0}&=&\sum_{k}\frac{ \hbar\Omega_k}{2\epsilon_0\epsilon_rV}\lvert R(\Omega_k)\rvert^2\cos(\Omega_k\tau),\nonumber \\
\end{eqnarray}
and its spectral representation, supposing a macroscopic crystal, can be calculated by integrating over the continuum of paraxial modes as
\begin{eqnarray}
\label{Geow}
\bra{0} \hat{G}_{eo}(\omega)\ket{0}&=&\frac{ \hbar \lvert\omega\rvert}{4\epsilon_0\sqrt{\epsilon_r} c S}\lvert R(\omega)\rvert^2,
\end{eqnarray}
which is the quantity measured in Ref. \cite{Benea-Chelmusarxiv}, while its frequency integral, corresponding to setting $\tau=0$ in \Eq{Geot}, was initially measured in Ref. \cite{Riek2015} using a single probe pulse.

\section{QVF in dispersive media}
We will now consider the case of a linear, local dielectric material characterised by an arbitrary dielectric function $\epsilon(\omega)$. The polaritonic formalism we will use can be extended to both lossy \cite{DeLiberato2017} and  inhomogeneous \cite{Gubbin2016} dielectrics, allowing to describe various resonator technologies \cite{Ballarini2019}, but in order to keep the notation as simple and clear as possible, we will focus here on a homogeneous and lossless material, with $\epsilon(\omega)$ symmetric and real over the whole real axis.
We notice that in Ref. \cite{DeLiberato2017} it was demonstrated that losses have anyway a limited impact on the structure of the ground state in linear dielectric systems.
Although our analytical results are derived for a generic dielectric function, for the sake of definiteness we will use as an example  the case of a single optically active Lorentz oscillator of frequency $\omega_x$ and vacuum Rabi frequency $g$
\begin{align}
\label{epsL}
\epsilon_L(\omega)&=\epsilon_r\left(1-\frac{4g^2}{\omega^2-\omega_x^2}\right).
\end{align}
\begin{figure}[htbp]
\begin{center}
\includegraphics[width=8.5cm]{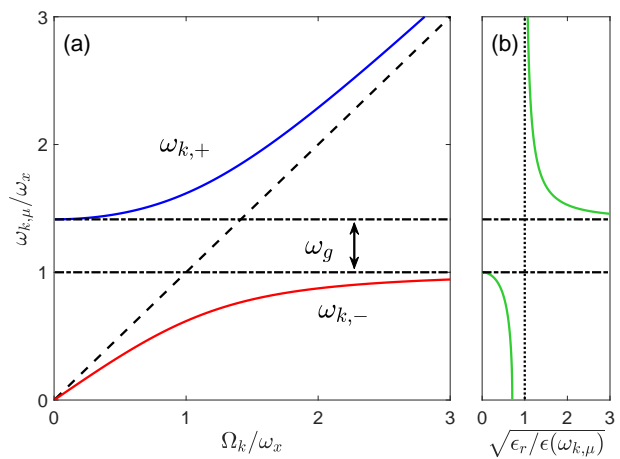}
\caption{(a) Dispersion of the upper (blue) and lower (red) polaritonic branches obtained with the Lorentz dielectric function in \Eq{epsL}, for $g=0.5\omega_x$.
The diagonal dashed black line denotes the bare photonic mode $\Omega_k$. (b) The green line represents the normalised result of the electro-optical measurement at the polaritonic frequencies, from \Eq{Rate}. The vertical dotted black line represents the uncoupled ($g=0$) value.
The polariton spectrum presents a gap $\omega_g$ between the bare frequency $\omega_x$ and the renormalised frequency $\sqrt{\omega_x^2+4g^2}$, shown as dash-dotted black lines.
\label{Fig_Dispersion}}
\end{center}
\end{figure}

Being the system linear its Hamiltonian can be diagonalised in terms of a set of polaritonic modes
\begin{align}
\label{Hp}
\hat{H}&=\sum_{k,\mu}\hbar\omega_{k,\mu} p^{\dagger}_{k,\mu}p_{k,\mu},
\end{align}
which satisfy bosonic commutation relations 
\begin{align}
\left[p_{k,\mu},p^{\dagger}_{k',\mu'}\right]=\delta_{k,k'}\delta_{\mu,\mu'},
\end{align}
where $\left[\cdot,\cdot\right]$ indicates the commutator. The index $\mu$ runs over all the polaritonic branches at a fixed wavevector, whose number depends on the exact form of $\epsilon(\omega)$. In the case of the Lorentz dielectric function in \Eq{epsL}, $\mu=\pm$ indexes the two polaritonic branches, whose dispersion $\omega_{k,\pm}$ are shown in Fig. \ref{Fig_Dispersion}(a). We notice here that the coupling opens a polaritonic gap in the system spectrum, where no propagative modes exist.

The bare photonic operators can then be written as linear superpositions of the polaritonic ones  
\begin{eqnarray}
\label{adef}
\hat{a}_{k}&=&\sum_{\mu}\left[\bar{X}_{k,\mu}\hat{p}_{k,\mu}-
Z_{k,\mu}\hat{p}_{k,\mu}^{\dagger}\right],
\end{eqnarray}
where the bosonicity  of the polariton and photon operators imposes the normalization condition on the Hopfield coefficients 
\begin{eqnarray}
\label{boscon}
\sum_{\mu}\left( \lvert X_{k,\mu}\rvert^2- \lvert Z_{k,\mu}\rvert^2\right)&=&1,
\end{eqnarray}
which also obey gauge invariance conditions
\begin{eqnarray}
\label{gaucon}
\Omega_k\left( {X}_{k,\mu}+{Z}_{k,\mu} \right)=\omega_{k,\mu}\left( {X}_{k,\mu}-{Z}_{k,\mu} \right).
\end{eqnarray}
The operator describing the $y$-polarised component of the paraxial electromagnetic field can now be written in terms of the polaritonic operators as
\begin{align}
\label{E}
E_v(t)&=\sum_{k,\mu}\sqrt{\frac{-\hbar\Omega_k}{2\epsilon_0V}} 
\left[\left(\bar{X}_{k,\mu}+\bar{Z}_{k,\mu} \right)\hat{p}_{k,\mu}e^{-i\omega_{k,\mu}t} -\text{h.c.}\right].
\end{align}
From \Eq{E}, performing flux quantization, we can read directly the polaritonic group velocity \cite{Huttner1991}
\begin{eqnarray}
\label{vg}
v^g_{k,\mu}=\frac{d \omega_{k,\mu}}{d k}=c\epsilon(\omega_{k,\mu})\left({X}_{k,\mu}+{Z}_{k,\mu}\right)^2.
\end{eqnarray} 

As clearly shown by the theory of open quantum systems in the ultrastrong coupling regime \cite{DeLiberato2009,Beaudoin2011,Bamba2014a}, polaritons  probe the electromagnetic environment at their own frequency.
Plugging \Eq{adef} and \Eq{vg} in \Eq{Seo}, the electro-optical operator can then be written in terms of polaritonic modes
\begin{align}
\label{SeoP}
\hat{S}_{eo}(t)&=\sum_{k,\mu}\sqrt{\frac{C\hbar\Omega_k v^g_{k,\mu}}{2\epsilon_0\epsilon(\omega_{k,\mu})cV}} 
\left[\hat{p}_{k,\mu}R(\omega_{k,\mu})e^{-i\omega_{k,\mu}t} -\text{h.c.}\right].
\end{align}
The time-resolved correlation function in the coupled polaritonic ground state $\hat{p}_{k,\mu}\ket{P}=0$ thus reads
\begin{align}
\label{GeoPt}
\bra{P} \hat{G}_{eo}(\tau)\ket{P}&=\sum_{k,\mu}\frac{\hbar \Omega_kv^g_{k,\mu}}{2\epsilon_0\epsilon(\omega_{k,\mu})cV} 
\lvert R(\omega_{k,\mu})\rvert^2\cos(\omega_{k,\mu}\tau).
\end{align}
We finally obtain the general expression for the spectral components of the correlation function
\begin{align}
\label{GeoPw}
\bra{P} \hat{G}_{eo}(\omega)\ket{P}&=\sum_{k,\mu}\frac{\pi\hbar \Omega_kv^g_{k,\mu}}{2\epsilon_0\epsilon(\omega_{k,\mu})cV} 
\lvert R(\omega_{k,\mu})\rvert^2\nonumber\\ &\times\left[\delta(\omega-\omega_{k,\mu})+\delta(\omega+\omega_{k,\mu}) \right].
\end{align}
In the case of a macroscopic crystal we can transform the sum over the paraxial modes in \Eq{GeoPw} and perform the integral, leading to
\begin{align}
\label{GeoPwc}
\bra{P} \hat{G}_{eo}(\omega)\ket{P}&=\frac{\hbar \lvert\omega\rvert}{4\epsilon_0\sqrt{\epsilon(\omega)}cS}
\lvert R(\omega)\rvert^2 I\left(\omega \right),
\end{align}
where $I\left(\cdot \right)$ is the indicator function over the polaritonic spectrum, equal to zero at the frequencies in which the polaritonic spectrum is gapped.  In the absence of propagative modes, the expected intensity of the QVF  vanishes.
Comparing \Eq{GeoPwc} to \Eq{Geow} we realise they are in the same form, once the proper dispersive dielectric function from $\epsilon(\omega)$ is used
\begin{align}
\label{Rate}
\frac{\bra{P} \hat{G}_{eo}(\omega)\ket{P}}{\bra{0} \hat{G}_{eo}(\omega)\ket{0}}&=\sqrt{\frac{\epsilon_r}{\epsilon(\omega)}}I(\omega).
\end{align}
A spectrally-resolved measurement of QVF through electro-optical sampling, once normalised over the vacuum value,  thus provides the frequency-dependent dielectric function of the material, effectively implementing an ellipsometry measurement over the quantum vacuum. 
Note that at the frequencies $\omega$ at which polariton modes exist and thus $I(\omega)=1$, the system admits propagative solutions, $\epsilon(\omega)>0$, and the square root in \Eq{Rate} is real.
Equivalent conclusions can be drawn in the case of a discrete spectrum, even though in this case \Eq{Rate} is not well defined, due to the different frequencies of the discrete modes in vacuum and in the dielectric. In Fig. \ref{Fig_Dispersion}(b) we plot the quantity in the right hand side of \Eq{Rate} for the Lorentz dielectric function in \Eq{epsL}.

\section{Discussion}
We have demonstrated that a measure of the time-correlations of QVF, once normalised over the uncoupled free-space vacuum result, provides us access to the same set of observables which we can measure with a linear optical characterisation of the sample. 
On one hand this result demonstrates electro-optical sampling of QVF is a  useful spectroscopic tool. It allows us to perform ellipsometry in the quantum vacuum, measuring the spectrally-resolved dielectric function without requiring any incoming photon in the probed frequency range.

On the other hand though, it raises some questions on the possibility of using electro-optical sampling as a direct experimental test of the ground state virtual photon population. 
The virtual photon population $N_k$, defined as the number of photons in the bare mode $\hat{a}_{k}$ emitted by the system after a non-adiabatic switch-off of the coupling \cite{Ciuti2005,DeLiberato2007} reads
\begin{eqnarray}
\label{Nk}
N_k=\bra{P}a^{\dagger}_{k}a_{k} \ket{P}&=&\sum_{\mu}\lvert {Z}_{k,\mu} \rvert^2.
\end{eqnarray}
Using \Eq{boscon} and \Eq{gaucon}, we can link $N_k$ to quantities which can be measured through electro-optical sampling of QVF through the evocative formula
\begin{eqnarray}
\label{partition}
\sum_{\mu} \frac{v^g_{k,\mu}}{4c} \left[1+\frac{1}{\epsilon(\omega_{k,\mu})} \right]&=&N_k+\frac{1}{2}.
\end{eqnarray}
The two sides of \Eq{partition} describe different physical measurements performed at different frequencies. The left hand side in fact relates to fluctuations inside the coupled system, measured through electro-optical sampling at the polaritonic frequencies $\omega_{k,\mu}$.
The right hand side describes instead photons emitted by the now uncoupled system and measured with any spectrally-resolved detector at the bare frequency $\Omega_k$.

Although it could seem that \Eq{partition}  allows indeed to measure $N_k$ via QVF electro-optical sampling, a few remarks are necessary.
The first is that the left hand side of \Eq{partition} contains quantities which can also be measured by standard linear spectroscopic techniques. Even though the QVF measurement can be said to actually probe the vacuum field, it cannot thus provide us access to observables qualitatively different from those we can acquire through a linear-optical characterisation of the dielectric.
The second is that there is no direct proportionality between the measured electro-optical correlations and the vacuum photon population. In order to write \Eq{partition} we need to rely on our theoretical modeling, linking the two quadratures of the field through \Eq{gaucon}.
This is to be expected given that the measure of a single field quadrature is not equivalent to a measure of the field population. Whether the magnetic quadrature of the QVF can be directly measured, obviating to this problem, remains an open question.
The lack of direct proportionality can also be verified from the fact that in Ref. \cite{DeLiberato2017} it is shown that for a medium described by \Eq{epsL}, $N_k$ doesn't present any resonant behaviour. This is clearly at odd with the results in Fig. \ref{Fig_Dispersion}(b), where a resonant behaviour can be observed in an interval of the order of the vacuum Rabi frequency $g$ around the polariton gap.

\section{Conclusions}
In this paper we developed the theory of spectrally-resolved electro-optical sampling of QVF in arbitrary linear, local dielectric materials. We demonstrated that such an approach allows us to implement a full linear optical characterisation, measuring the frequency-dependent dielectric function. This proves its usefulness as an alternative spectroscopic tool for the characterisation of linear dielectrics. Its ability to perform investigations on the properties of the interacting quantum vacuum fundamentally different from those implementable with linear optical techniques remains nevertheless unclear.  

\section{Acknowledgements}
I acknowledge useful feedback and discussions with I.-C. Benea-Chelmus, J. Faist, N. Lambert, and F. F. Settembrini.
I am a Royal Society Research Fellow and I acknowledge support from the Innovation Fund of the EPSRC Programme EP/M009122/1 and from the Philip Leverhulme prize of the Leverhulme Trust.

\bibliography{References}

\end{document}